\documentclass[twocolumn,showkeys,preprintnumbers,amsmath,amssymb]{revtex4}
\usepackage{graphicx}
\usepackage{bm}
\usepackage{graphicx}

\usepackage{psfrag}  
\usepackage{amsmath,amsthm}
\usepackage{amsfonts}
\usepackage{amssymb}
\usepackage{bm}

\begin{document}


\title{Topological  reversibility and causality in  feed-forward  networks}  
\author{Bernat  Corominas-Murtra$^1$,  Carlos Rodr\'iguez-Caso$^1$,  Joaquin
  Go\~ni$^2$ and Ricard V. Sol\'e$^{1,3,4}$}

\affiliation{$^1$ ICREA-Complex Systems  Lab, Universitat Pompeu Fabra
  (Parc de Recerca Biom\`edica de Barcelona). Dr Aiguader 88, 08003 Barcelona, Spain\\ 
  $^2$Functional
  Neuroimaging  Lab. Department of  Neurosciences. Center  for Applied
  Medical Research. University of Navarra. Pamplona, Spain\\
  $^3$Santa Fe
  Institute, 1399  Hyde Park Road, New Mexico  87501, USA\\
  $^4$Institut
  de Biologia Evolutiva. CSIC-UPF.  Passeig Mar\'itim de la Barceloneta,
  37-49, 08003 Barcelona, Spain.}

\begin{abstract}
Systems   whose organization  displays causal  asymmetry   
constraints, from evolutionary trees to  river basins or
 transport networks, can be often described in terms of directed 
 paths (causal flows) on a discrete state space. Such a set of paths
 defines a feed-forward, acyclic network.  A key problem associated with
these systems involves characterizing their intrinsic degree of path
reversibility: given an end node in the graph,  what is the uncertainty
of recovering the process backwards until the origin? Here we propose
a novel concept, {\em topological reversibility}, which rigorously
weigths such uncertainty in path dependency quantified  as the minimum  
amount of  information  required to
successfully  revert  a causal  path.  
Within the proposed framework we also analytically characterize limit cases for both topologically reversible  and maximally entropic structures. The relevance of these measures within the context of evolutionary dynamics is highlighted.
\end{abstract}

\maketitle

\maketitle
\section{Introduction}
\label{Introduction}
Causality is the fundamental  principle pervading dynamical
processes.  Any  set of time-correlated events, from  the development of
an organism to historical changes, defines a feed-forward structure of
causal  relations captured  by  a family  of  complex networks  called
directed acyclic graphs (DAGs). Their structure has recently attracted
the   interest    of   researchers   \cite{Lehmann:2003,   Csardi2007,
  Valverde2007, Karrer2009} since DAGs represent time-ordered
processes as well as a broad number of natural and artificial systems. Examples would
include simple electronic circuits \cite{Clay1971}, feed-forward neural
\cite{Haykin1999} and transmission networks \cite{Frank1972},
river basins \cite{Rodriguez-Iturbe1997}, or even some food webs and chemical structures \cite{Bonchev2005}.

A  paradigmatic example  of a  causal structure  is the  chart  of the
relations among states followed  by a computational process through
time.   Intimately  linked  to the  topology  of  the
computational  chart of  consecutive states,  a fundamental  feature of
computations   is   its   degree   of  {\em   logical   reversibility}
\cite{Landauer:1961, Bennett:1973}.  Indeed, it is said that a process
is  {\em  logically  reversible}   when,  if  reverting  the  flow  of
causality, i.e.  going backwards from the computational outputs to
their inputs, we can unambiguously recover the causal structure
of the process.  Roughly speaking,  if we have a computer performing a
function   $g:\mathbb{N}\to\mathbb{N}$   and   we  can   unambiguously
determine the input $u$ from the only knowledge of the value $v=g(u)$,
we say that the function  is logically reversible. Otherwise, if there
is uncertainty in  determining $u$ from the only  knowledge of $v$, we
say  that the  function  is {\em  logically  irreversible}, and  thus,
additional information  is needed to successfully  reconstruct a given
computational path.

Analogously, the potential scenarios emerging from an evolutionary process raise similar questions. Within evolutionary biology, a relevant problem is how predictable is evolutionary dynamics. In particular, it has been asked what would be the result of going backwards and "re-playing the tape of evolution" \cite{Gould1990, Fontana1994}. Since this question pervades the problem of how uncertain or predictable is a given evolutionary path, it seems desirable to actually provide a foundational framework.  


In  this paper,  we  analytically  extend  the concept  of
logical  reversibility  to the study of any  causal  structure  having  no  cyclic
topologies,  thereby  defining a  broader  concept  to  be named  {\em
  topological  reversibility}.  Whereas  thermodynamical irreversibility
implies   thermodynamical entropy   production   \cite{DeGroot:1963,
  Lebon:2008}, topological irreversibility implies statistical entropy
{\em production}.
In general, we  will say that a DAG  is {\em topologically reversible}
if  we can  unambiguously  recover  a path  going  backwards from  any
element to  the origin.  Genealogies  and phylogenies are  examples of
tree-like structures  where  a chronological order can  be established
among the events and an unambiguous reconstruction of the lineage can
be performed for every element of the graph \cite{Schuster2010}.  Following this argument,
we will  label a graph  as {\em topologically irreversible}  when some
uncertainty is  observed in  the reconstruction of  trajectories.

As shown below, the  entropy presented here weigths  the extra  amount of information that would be required to recover the causal flow backwards.   Information  measures are not
new in the study of complex networks \cite{Schneidman:2003, Sole:2004,
  Dehmer:2008,  Dehmer:2008b,  Bianconi:2009, Estrada:2009, Ulanowicz1986},  although
such     measures    accounted    for     connectivity    correlations
\cite{Schneidman:2003,  Sole:2004, Dehmer:2008, Dehmer:2008b}  or were
used to characterize  a  Gibbsian formulation  of the
statistical  mechanics of  complex networks  \cite{Bianconi:2009}.  We finally note that the starting point of our formalism resembles the
classical theory  of Bayesian  networks.  However, the  particular  treatment of reversibility proposed  here is
qualitatively different  from the concept of  uncertainty  used in such a framework and   closer  to   the  one  described  in
\cite{Estrada:2009}.

The     paper    is    organized     as    follows:     In    section
\ref{TheoreticalBackground} we  provide the basic  concepts underlying
our analytical derivations.   Section  \ref{FF1}  provides the  general  mathematical
definition  of  topological  reversibility  and the  general
expression for the average  uncertainty associated to the reversion of
the causal flow. This  is consistently derived from
the properties of the adjacency matrix.
In  section \ref{Extremal}  we  consider two limit cases, finding
 the  exact  analytic form for  their  entropies and predicting the  uncertain configuration.   Finally, in section \ref{HierarchyIrreversibility}
we  outline the generality  and relevance  of our  results in  terms of
characterizing  DAG structure.
\begin{figure*}
\begin{center}
\includegraphics[width=16cm]{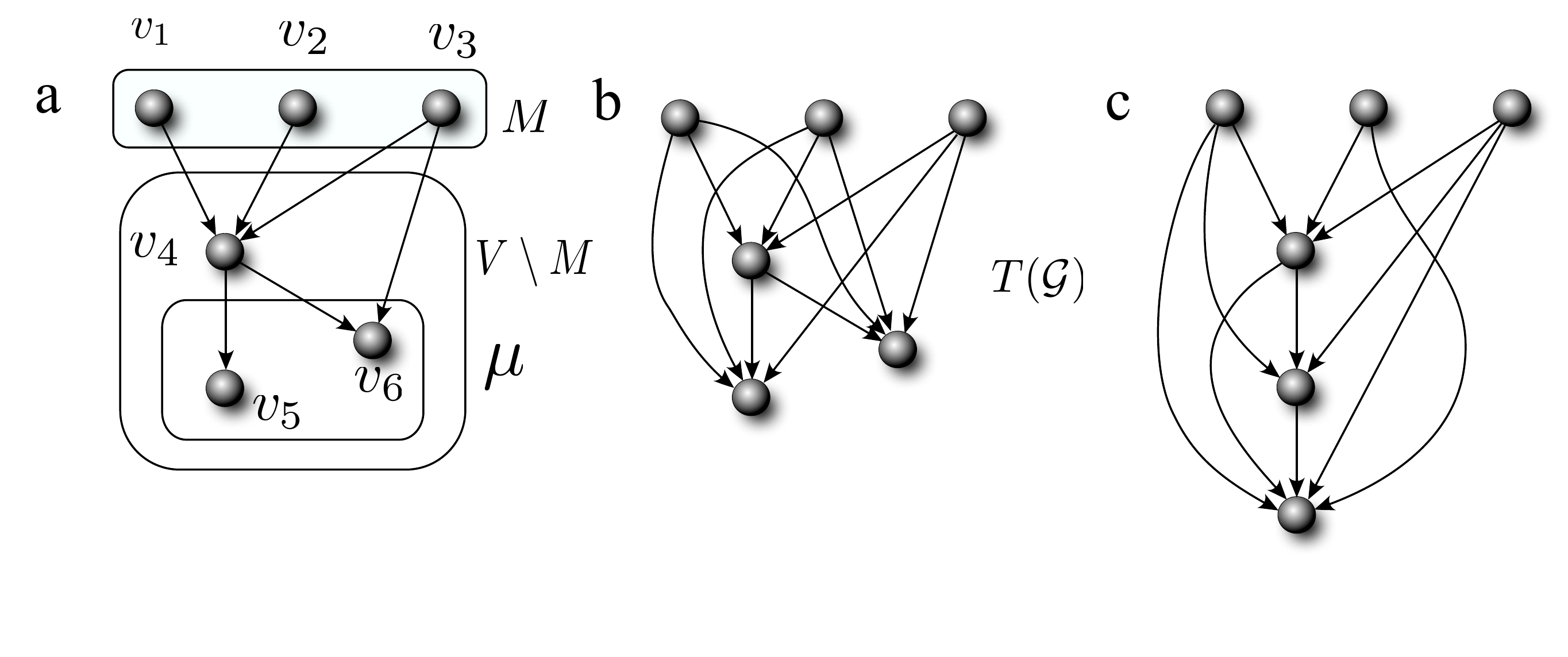}
\caption{Some illustrative DAGs. A topologically irreversible DAG ${\cal G}(V,E)$, where $M$ denotes the set of maximals, 
 $\mu$ the set of minimals and the $V\setminus M$ set the set of non-maximals (a). The respective transitive 
 closure, ${\cal T}({\cal G})$ is shown in (b), A linear ordering of the set $V\setminus M$ of ${\cal G}(V,E)$ is displayed in (c) where any node of the maximal set is
  connected to any  node of the set $V\setminus M$. This is an special structure displaying maximal entropy (see text).}
\label{fig1}
\end{center}
\end{figure*}

\section{Theoretical Background}
\label{TheoreticalBackground}
The theoretical roots  of this paper stem from  fundamental notions of
directed  graph theory  \cite{Gross:1998, Bollobas:1998},  ordered set
theory   \cite{Kelley:1955,   Suppes:1960}   and  information   theory
\cite{Shannon:1948, Khinchin:1957, Ash:1990, Thomas:2001}.  Specifically, we make use
of Shannon's  entropy which, as originally defined, quantifies
the  uncertainty associated  to certain  collections of  random events
\cite{Khinchin:1957,  Ash:1990}.  In  our framework,  the entropy  in a
given feed-forward  graph measures the  uncertainty in reversing  the causal
flow  depicted by  the  arrows\footnote{It  is  important to  notice that  the
results  reported in  this  paper  are independent  on  the number  of
connected components  displayed by the  DAG. However, we  will tacitly
assume  that  one single  connected  component  linking  all nodes  is
present,  unless the  contrary  is indicated.  An intuitive  statement
guides our choice: Two unconnected components are causally independent
and, therefore, they must be treated as independent entities.}.

\subsection{Directed graphs and orderings}
\label{BasicGDef}

Let    ${\cal    G}(V,    E)$    be   a    directed    graph,    being
$V=\{v_1,...,v_n\},\;|V|=n$, the  set of nodes,  and $E=\{\langle v_k,
v_i\rangle, ...,  \langle v_j, v_l\rangle\}$  the set of  edges -where
the order, $\langle v_k, v_i\rangle$ implies that there is an arrow in
the following  direction: $v_k\rightarrow v_i$.  Given  a node $v_i\in
V$, the number of outgoing links, to be written as $k_{out}(v_i)$, is called
the {\em out-degree} of $v_i$ and the number of ingoing links of $v_i$
is called the {\em in-degree} of $v_i$, written as $k_{in}(v_i)$. The {\em
  adjacency  matrix} of  a given  graph ${\cal  G}$, $\mathbf{A}({\cal G})$ is defined as $A_{ij}({\cal G})= 1 
\leftrightarrow \langle   v_i,  v_j\rangle\in   E$; and  $A_{ij}({\cal G})= 0$ otherwise.
 Through the adjacency matrix, $k_{in}$ and $k_{out}$ are computed as
\begin{equation}
k_{in}(v_i)=\sum_{j\leq n}A_{ji}({\cal G});\;\;\;\;k_{out}(v_i)=\sum_{j\leq n}A_{ij}({\cal G}).
\label{kin}
\end{equation}
Furthermore,  we will use the  known  relation between  the
$k$-th  power of  the adjacency  matrix and  the number  of  paths of
length  $k$ going  from  a given  node  $v_i$ to  a  given node  $v_j$
Specifically,
\begin{equation}
(\mathbf{A}({\cal  G}))_{ij}^k=(\overbrace{\mathbf{A}({\cal  G})\times...
\times \mathbf{A}({\cal  G})}^{k\;{\rm times}})_{ij}\nonumber
\end{equation}
is the  number of paths  of length $k$  going from node $v_i$  to node $v_j$ \cite{Gross:1998}.

A {\em  feed-forward} or  {\em directed acyclic  graph} is  a directed
graph  characterized by  the absence  of cycles:  If there  is  a {\em
  directed path} from $v_i$ to $v_k$ (i.e., there is a finite sequence
 $\langle v_i, v_j\rangle,  \langle v_j,  v_l\rangle,\langle v_l,
v_s\rangle, ...,  \langle v_m,  v_k\rangle \in E$)  then, there  is no
directed path from $v_k$ to $v_i$. Conversely, the matrix $\mathbf{A}^T({\cal
  G})$ depicts a DAG with the same underlying structure but having all
the arrows  (and thus,  the causal flow)  inverted. Given  its acyclic
nature, one can find a finite value $L({\cal G})$ as follows:
\begin{equation}
L({\cal G})=\max\{k:(\exists v_i, v_j\in V:(\mathbf{A}({\cal G}))^k_{ij}\neq 0)\}.
\label{K}
\end{equation}
It is  easy to see that $L({\cal  G})$ is the length  of the
longest path of the graph.  The existence of such $L({\cal G})$ can be
seen  as a  test for  acyclicity.   However, the  use of  leaf-removal
algorithms \cite{Lagomarsino2006, Rodriguez-Caso2009},  i.e. the  iterative  pruning of  nodes without  outgoing
links, is  by far  more suitable  than the above  method, in  terms of
computational costs.  In a
DAG, a leaf-removal algorithm removes completely the graph in a finite
number of iterations, specifically, in $L({\cal G})$ iterations -see eq. (\ref{K}).

Now we study the interplay between DAGs and order relations. Borrowing
concepts from order theory \cite{Suppes:1960}, we define the following
set:
\begin{equation}
M=\{v_i\in V:k_{in}(v_i)=0\},
\label{MaximalSet}
\end{equation}
to be  named the set  of {\em maximal  nodes} of ${\cal G}$,  by which
$|M|=m$.  The  set of all  paths $\pi_1,...,\pi_s$, $s\geq  |E|$, from
$M$ to a  given node $v_i\in V\setminus M$  is indicated as $\Pi({\cal
  G})$.  Given a node $v_i\in V\setminus M$, the set of all paths from
$M$ to $v_i$ is written as $\Pi(v_i)\subseteq \Pi({\cal G})$.  Furthermore, we will define the set
$v(\pi_k)$ as the set of  all nodes participating in this path, except
the  maximal one.   Additionally,  one  can define  the  set of  nodes
with $k_{out}=0$  as the  set of {\em  minimal nodes}  of ${\cal
  G}$, to be  named $\mu$.  Notice that the  absence of cycles implies
that $m\geq 1$ and that the set of minimals $\mu$ must also contain at
least one element -see fig. (\ref{fig1}a).

Attending to the  node relations depicted by the  arrows, and due to the acyclic property, at least one
node  ordering can  be defined,  establishing a  natural  link between
order theory  and DAGs.   This order is  achieved by labeling  all the
nodes with  sequential natural  numbers and obtaining  a configuration
such that:
\begin{equation}
(\forall \langle v_i, v_j \rangle \in E)(i<j).
\label{i>j}
\end{equation}
Accordingly,   DAGs   are   {\em   ordered   graphs}
\cite{Karrer2009}.    However,   as   order   relations   imply   {\em
  transitivity}, it  is not  the DAG but  its transitive  closure what
properly defines  the order  relation among the  elements of  $V$. The
{\em  transitive closure}  of ${\cal  G}$ (see fig. \ref{fig1}b), to  be written  as $T({\cal
  G})=(V_T,  E_T)$ is  defined as  follows:  Any pair  of nodes  $v_i,
v_k\in V$  by which  there is at  least one  path going from  $v_i$ to
$v_k$  are  connected through  a  link  $\langle  v_i, v_k\rangle$  in
$T({\cal  G})$.  In  this framework,  for  a given  number of  maximal
nodes, in the transitive closure the addition of a link either creates
a cycle or destroys a maximal  or minimal node.  If the pairs defining
the set of  links of $T({\cal G})$ are conceived as  the elements of a
set  relation $E_T\subset V\times  V$, such  a relation  satisfies the
following three properties:
\begin{eqnarray}
&i)&\nexists \langle v_k, v_k \rangle,\nonumber\\
&ii)&(\langle v_i, v_k\rangle \in E_T)\Rightarrow (\langle v_k, v_i\rangle \notin E_T),\nonumber\\
&iii)&(\langle v_i, v_k\rangle \in E_T \wedge \langle v_k, v_j \rangle \in E_T)\Rightarrow (\langle v_i, v_j \rangle \in E_T).\nonumber
\end{eqnarray}
The DAG definition  implies that $E$ directly satisfies  the two first
conditions whilst  the third one (transitivity) is  only warranted for
$E_T$.   Thus, only  $E_T$ holds  all  requirements  to  be an  {\em  order
  relation},  specifically,   a  {\em  strict   partial  order}.   The
transitive closure  of a  given DAG  can be obtained  by means  of the
so-called {\em Warshall's} algorithm \cite{Gross:1998}.

Finally, a  subgraph ${\cal F}(V_{\cal F},  E_{\cal F})\subseteq {\cal
  G}$ is  said to be {\em  linearly ordered} or  {\em totally ordered}
provided that for all pairs of nodes $v_i,v_k\in V_{\cal F}$ such that
$k<i$, then
\begin{equation}
\langle v_k,v_i\rangle\in E_{\cal  F}.
\label{Linear}
\end{equation}
Let us notice that  if  we understand  $E_{\cal  F}$ as  a set  relation
$E_{\cal  F}\subset V_{\cal F}\times  V_{\cal F}$,  $E_{\cal F}$  is a
{\em  strict linear  order}. If  ${\cal  G}$ is  linearly ordered  and
${\cal  W}\subset  {\cal  G}$,  we  refer  to ${\cal  G}$  as  a  {\em
  topological sort} of ${\cal W}$ \cite{Gross:1998}.


\subsection{Uncertainty}
\label{SecUncertainty}
According  to  classical information  theory  \cite{Shannon:1948, Khinchin:1957, Ash:1990, Thomas:2001}, let  us
consider a system $S$ with  $n$ possible states, whose occurrences are
governed by a random variable $X$ with an associated probability mass function formed by $p_1,
...,   p_n$.    According   to   the   standard   formalization,   the
\textit{uncertainty} or {\em entropy} associated to $X$, to be written
as $H(X)$, is:
\begin{equation}
H(X)=-\sum_{i\leq n}p_i\log p_i,
\label{DefEntrop}
\end{equation}
which  is  actually  an average  of
$\log(1/p(X))$ among all events of $S$, namely, $H(X)=\left\langle \log(1/p(X))\right\rangle$,
where  $\langle...\rangle$ is the  {\em expectation}  or average  of the
random  quantity  between parentheses.   As  a  concave function,  the
entropy   satisfies   the    so-called   {\em   Jensen's   inequality}
\cite{Thomas:2001}, which reads:
\begin{equation}
\left\langle \log\frac{1}{p(X)}\right\rangle\leq \log\left\langle\frac{1}{p(X)}\right\rangle\leq \log n,
\label{Jensen}
\end{equation}
The  maximum  value $\log  n$  is  achieved  for 
$p_i=1/n$ for all $i=(1,...,n)$.  Jensen's inequality provides an upper bound on the entropy
that will  be used below. Analogously, we  can define
the {\em  conditional entropy}.  Given another  system $S'$ containing
$n'$ values or choices, whose  behavior is governed by a random variable
$Y$,  let  $\mathbb{P}(s'_i|s_j)$ be  the  conditional probability  of
obtaining $Y=s'_i\in  S'$ if we  already know $X=s_j\in S$.  Then, the
conditional  entropy of $Y$  from $X$,  to be  written as  $H(Y|X)$, is
defined as:
\begin{equation}
H(Y|X)=-\sum_{j\leq n}p_j\sum_{i\leq n'}\mathbb{P}(s'_i|s_j)\log \mathbb{P}(s'_i|s_j).
\label{H(Y|X)}
\end{equation}
which is typically  interpreted   as  a   noise  term  in   information  theory. Such a noise term can be interpreted as the minimum amount of extra bits
needed to unambiguously determine the input set from the only knowledge of the output set. This will be the key quantity of our paper, for it accounts for the {\em dissipation} of information in a given process.

\section{Topological reversibility and entropy}
\label{FF1}
\begin{figure}
\begin{center}
\includegraphics[width=10cm]{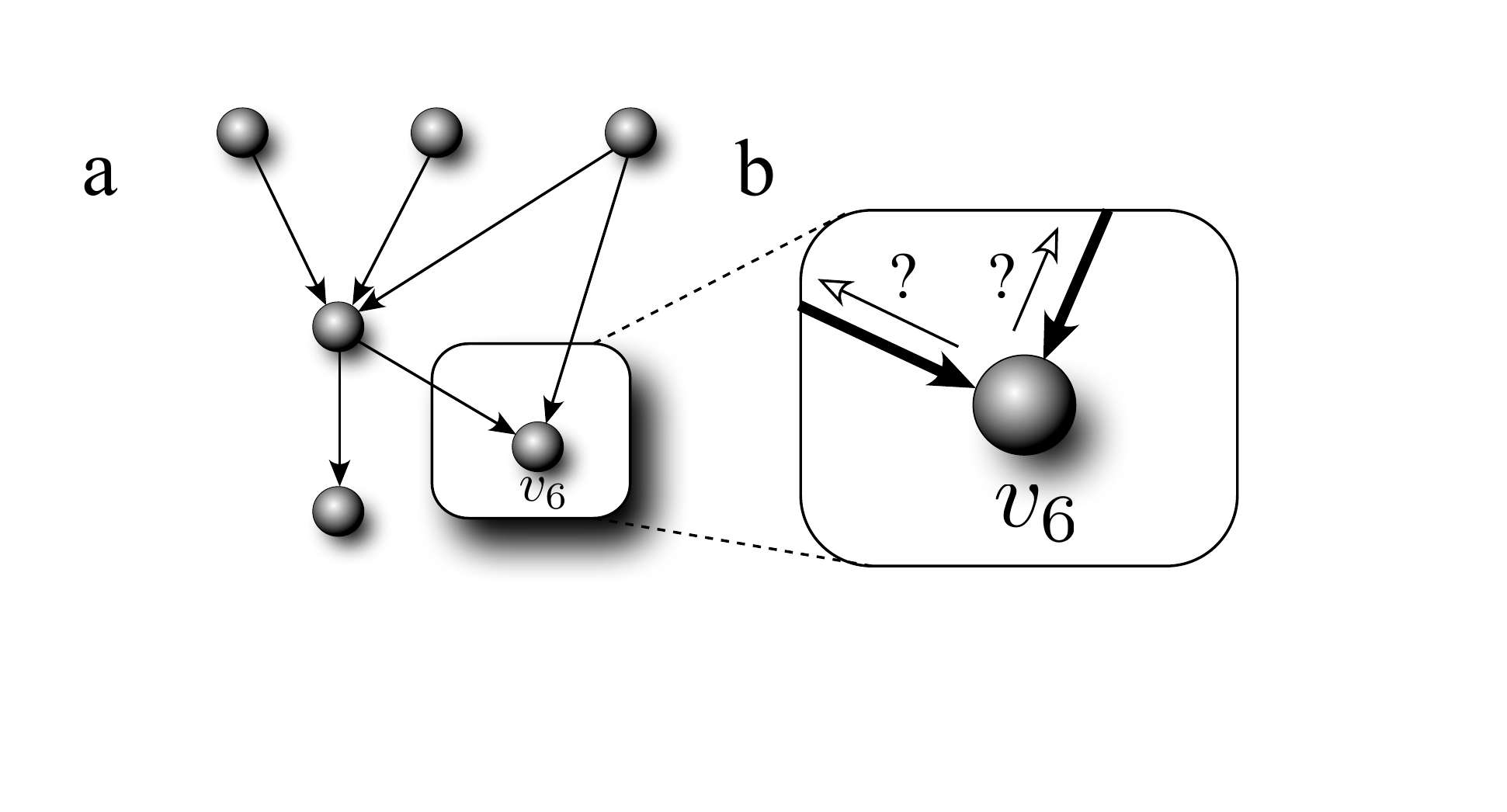}
\caption{Uncertainty in the reversal of causal flows in a DAG. Notice that more than a pathway, with more or less probability to be chosen, connect maximals  from each terminal (a). Given a node  ($v_6$) receiving two inputs, we consider two different alternatives to go backwards. The uncertainty in this particular case is obtained by computing $h_L({v_i})$ from eq. (\ref{h_L}), i.e., $h_L(v_6)=\log{2}$ assuming equiprobability in the selection  (b)}.   
\label{fig2}
\end{center}
\end{figure}
Let us  imagine that a node $v_i\in V\setminus M$  of a given
DAG ${\cal G}$,  receives the visit of a  random walker that follows the flow chart depicted by the DAG. We only
know  that it  began  its walk  at a given  maximal  node and it followed  a
downstream random  path attending to  the directions of the  arrows to
reach the node $v_i$.   Suppose also that 
the  global  structure of  the  graph is unknown.   What  is the  uncertainty 
associated to the followed path?  In other words,  what is the amount
of  information  we need,  on  average,  to  successfully perform  the
backward process?

\subsection{The definition of entropy}
As we mentioned above, the  starting  point of  our  derivation  is  close to  treatment of Bayesian networks  \cite{Jensen:2001}. In our approach, the
first  task  is to  define  the probability  to  follow  a given  path
$\pi_k\in \Pi(v_i)$ when reverting the process.  Let $v(\pi_k)$ be the
set of nodes participating in the path $\pi_k$ except the maximal ones. Maximal nodes are not included in this set because they are the ends of the path of the reversal process.
The  probability to  chose such  a path  from node  $v_i$ by  making a
random decision at every crossing  when reverting the causal flow will
be:
\begin{equation}
\mathbb{P}(\pi_k|v_i)=\prod_{v_i\in v(\pi_k)}\frac{1}{k_{in}(v_j)}.
\label{Pprod}
\end{equation}
Consistently:
\begin{equation}
\sum_{\pi_k\in\Pi(v_i)}\left(\prod_{v_j\in v(\pi_k)}\frac{1}{k_{in}(v_j)}\right)=1.\nonumber
\end{equation}
As  $\mathbb{P}$ is  a probability  distribution, we  can  compute the
uncertainty associated to a reversal  of the causal flow, starting the
reversion  process from  a given  node  $v_i\in V\setminus  M$, to  be
written as $h(v_i)$:
\begin{eqnarray}
h(v_i)=-\sum_{\pi_k\in\Pi (v_i)}\mathbb{P}(\pi_k|v_i)\log\mathbb{P}(\pi_k|v_i)
\end{eqnarray}
The overall  uncertainty of ${\cal  G}$, written as $H({\cal  G})$, is
computed by averaging $h$ over all non-maximal nodes, i.e:
\begin{eqnarray}
H({\cal G})&=&-\sum_{v_i \in V\setminus M}p(v_i)
\sum_{\pi_k\in\Pi (v_i)}\mathbb{P}(\pi_k|v_i)\log\mathbb{P}(\pi_k|v_i)\nonumber\\
&=&\sum_{v_i \in V\setminus M}p(v_i) h(v_i).
\label{Entropia1}
\end{eqnarray}

\subsection{The transition matrix $\Phi$ and its relation to the adjacency matrix}
The main  combinatorial object  of our approach  is not  the adjacency
matrix   but
instead a  mathematical representation of  the probability to  visit a
node $v_i\in V\setminus M $  starting the backward flow
from  a given,  different node  $v_k\in V\setminus  M$  regardless the
distance  separating  them.   As  we  shall  see,  this  combinatorial
information can  be encoded in a  matrix, to be  named {\em transition
  matrix} $\Phi$ and  we  can explicitly obtain  it
from $\mathbf{A}({\cal G})$. We begin by defining
\begin{equation}
  V(\Pi(v_j))\equiv \bigcup_{\pi_k\in\Pi(v_j)}v(\pi_k),
\end{equation}
and we can  see that:
\begin{eqnarray}
h(v_i)&=&-\sum_{\pi_k\in\Pi (v_i)}\mathbb{P}(\pi_k|v_i)\log\mathbb{P}(\pi_k|v_i)\nonumber\\
&=& \sum_{\pi_k\in\Pi (v_i)}\left[\sum_{v_j\in v(\pi_k)}\mathbb{P}(\pi_k|v_i)\log (k_{in}(v_j))\right]\nonumber\\
&=& \sum_{v_j\in V(\Pi_i)}\log (k_{in}(v_j))\left[\sum_{\pi_k:v_j\in v(\pi_k)}\mathbb{P}(\pi_k|v_i)\right]\nonumber\\
&=&\sum_{v_k\in V\setminus M}\phi_{ik}({\cal G})h_L(v_k).
\label{h(v_i)}
\end{eqnarray}
Let  us explain eq.   (\ref{h(v_i)}) and  its consequences.   First we
define $h_L(v_i)$ as:
\begin{equation}
h_L(v_i)=\log (k_{in}(v_i)). \label{h_L},
\end{equation}
where $L$ indicates the
amount of {\em local} entropy introduced in a given node when performing the
reversion process -see fig (\ref{fig2}). Thereby, it is  the amount of information needed to
properly  revert the  flow backwards when  a bifurcation  point is reached having
$k_{in}$    possible    choices. Secondly,  we define $\phi_{ik}$  as the coefficients  of a
$(n-m)\times  (n-m)$  matrix  $\Phi({\cal G})=[\phi_{ik}({\cal  G})]$,
i.e. our {\em transition matrix} ${\cal G}$:
\begin{eqnarray}
\phi_{ij}({\cal G})=\sum_{\pi_k:v_j\in v(\pi_k)}\mathbb{P}(\pi_k|v_i).\nonumber
\end{eqnarray}
This represents the probability  to  reach $v_j$  starting  from $v_i$.  
Now we derive the general expression for $\Phi$. The derivation allows us  to  obtain  a  consistent mathematical  definition  of  the
transition matrix in  terms of $\mathbf{A}({\cal G})$.   We first notice
two  important facts linking  paths  and the  powers of  the adjacency
matrix that are only generically valid in DAG-like networks. First, we
observe that:
\begin{equation}
|\Pi (v_i)|=\sum_{j\leq L({\cal G})}\sum_{l:v_l\in M}(\mathbf{A}^T({\cal G}))^j_{il},
\label{Pivi}
\end{equation}
being $L({\cal  G})$ the length  of the longest path  of the graph as
defined  by (\ref{K}).   Analogously,  the  number  of  paths  of
$\Pi(v_i)$ crossing $v_k$, to be written as $\alpha_{ik}$ is:
\begin{eqnarray}
\alpha_{ik}&\equiv& |\{\pi_j\in \Pi(v_i):v_k\in v_i(\pi_j) \}|\nonumber\\
&=& \sum_{j\leq L({\cal G})}\left(\mathbf{A}^T({\cal G})\right)^j_{ik}.
\label{alpha}
\end{eqnarray}
The  above quantities provide the  number of  paths. To
compute the  probability to reach a  given node, we have  to take into
account the probability to follow a given path containing such a node,
defined in  (\ref{Pprod}).   To rigorously connect it  to the
adjacency    matrix,   we   first    define   an    auxiliary, $(n-m)\times (n-m)$   matrix
$\mathbf{B}({\cal G})$, namely:
\begin{equation}
B({\cal G})_{ij}=\left(A_{ij}({\cal G})\right)\left(\sum_{j\leq n}A_{ij}({\cal G})\right)^{-1}=\frac{A_{ij}({\cal G})}{k_{in}(v_i)},
\label{B}
\end{equation}
where $v_i,v_j\in V\setminus M$.
From this definition, we obtain the explicit dependency of $\Phi$ from
the adjacency matrix, namely\footnote{We observe that  matrix $\mathbf{B}^T$ is the matrix corresponding to a Markov process \cite{VanKampen:2007} depicting a random walker walking against the flow},
\begin{equation}
\phi_{ij}({\cal G})=\sum_{k\leq L({\cal G})}\left(\mathbf{B}^T({\cal G})\right)^k_{ij}.
\label{phi(B)}
\end{equation}
and accordingly, we have
\begin{equation}
\phi_{ii}({\cal G})= \left(\mathbf{B}^T({\cal G})\right)^0_{ii}=1.
\end{equation}
It is worth to mention that $\Phi({\cal G})$  resembles the transition matrix
related to the concept of  {\em information mobility} \cite{Estrada:2009}.  In the general
case of non-directed  graphs, one can assume the  presence of paths of
arbitrary length, which  leads (using a correction factor tied to the length of the path)
up to an asymptotic form of  the transition matrix  in terms of
the exponential of the adjacency matrix.  However,
the intrinsic finite nature of the paths in a given DAG makes
the above asymptotic treatment non viable.

\subsection{The general form of the Entropy}

Let us now define the overall entropy in a compact form,
only depending  on the  adjacency matrix of  the graph.   From
eqs. (\ref{H(Y|X)},  \ref{Entropia1}, \ref{h(v_i)}), we
obtain 
\begin{equation}
H({\cal G})=\sum_{v_i\in V\setminus M}p(v_i)\sum_{v_k\in V\setminus M}\phi_{ik}({\cal G})h_L(v_k).
\label{HGeneral}
\end{equation}
This is the central equation of this paper.  This measure quantifies 
the additional information (other than topological one) to properly revert the causal flow. We observe that this expression is a \textit{noise} term within standard information theory \cite{Ash:1990}.
In this equation  we have been able to
decouple  the combinatorial  term  associated to  the multiplicity  of
paths  at one  hand, and  the particular  contribution to  the overall
uncertainty of every node, at the  other hand. The former is fulfilled  by the
matrix $\Phi$,  which encodes combinatorial properties  of the system, and how they
influence in the computation of the entropies.  The latter is obtained from the set of local entropies
$h_L(v_1),...,h_L(v_{n-m})$. These terms account for  the contribution of
local  topology -i.e. the uncertainty when choosing an incoming link at the node level in the reversion of the causal flow-  to  the  overall  entropy.   This
uncoupling is a consequence of the  extensive property of the entropy and,
putting  aside  its conceptual  interest,  simplifies all  derivations
related to the  uncertainties, since we are not  forced to compute the
complex  series arising  in the  brute-force calculation  of entropies.  
This general expression of the entropy can be simplified if we assume that $\forall v_i\in V\setminus M$, $p(v_i)=1/(n-m)$. Therefore, by defining 
\begin{equation}
Q({\cal G})=\sum_{v_i\in V\setminus M}\sum_{v_k\in V\setminus M}\phi_{ik}({\cal G})h_L(v_k)
\label{Q}
\end{equation}

and thus $H({\cal G})$ is expressed as:
\begin{equation}
H({\cal G})=\frac{1}{n-m}Q({\cal G})
\label{HFHL}
\end{equation}
Finally,  we recall
that the above entropy is bounded by Jensen's inequality
(\ref{Jensen}) i.e.,
\begin{equation}
H({\cal       G})\leq       \frac{1}{n-m}\sum_{v_i\in       V\setminus
  M}\log(|\Pi(v_i)|).
\label{hMG1}
\end{equation}
Notice that the quantity on the right side of eq.  (\ref{hMG1}) is the
uncertainty  obtained  by considering  all  paths  from  $M$ to  $v_i$
equally likely to occur.

\subsection{Topological reversibility}

Having  defined an  appropriate and  well grounded  entropy measure, now  we can
discuss the  meaning of  {\em topological (ir)reversibility}.   Let us
first make a qualitative link  with   standard  theory   of   irreversible
thermodynamics,  where irreversibility  is  tied to  the parameter  of
entropy  production  $\sigma^s$  in  the  entropy  balance
equation \cite{Lebon:2008}. Here,  $\sigma^s=0$ depicts  {\em thermodynamically
  reversible}    processes,    whereas    $\sigma^s>0$   appears    in {\em  irreversible}  processes  \cite{DeGroot:1963,
  Lebon:2008}.   Irreversibility  is rooted  in  the impossibility  of reverting the process without generating  a negative amount  of entropy,
which contradicts to the second  law of thermodynamics.
Consistently, we  will call  {\em topologically reversible}  those DAG
structures such  that
\begin{equation}
H({\cal G})=0.\nonumber
\end{equation}
In those structures (they belong to the set of trees, as we shall see
in  the following  section) no  ambiguity arises  when  performing the
reversion process. On the contrary, a given DAG by which
\begin{equation}
H({\cal G})>0\nonumber
\end{equation}
will be  referred to as {\em topologically  irreversible}. DAGs having
$H({\cal G})>0$  display some degree  of uncertainty taking the
causal flow backwards,  since the  reversion process is  subject to  some random inevitable
decisions.  In these cases, $H({\cal G})$ is the average of the amount
of  extra information  needed  to successfully  perform  the process backwards.
Similarly, the successful  reversion of a
thermodynamically irreversible process  would imply the (irreversible)
addition  of external  energy, or  that the  reversion of  a logically
irreversible  computation   requires  an  extra   amount  of  external
information to solve  the ambiguity arising in rewinding  the chain of
computations.   In this context, for example, reversible  computation is
defined  by  considering  a  system  of  storage  of  history  of  the
computational  process \cite{Bennett:1973}.   Furthermore,  we observe
that, roughly speaking, we  can associate the logical (ir)reversibility of
a computational process to  the topological (ir)reversibility of its
DAG  representation. In  our  study, the  adjective {\em  topological}
arises  from the  fact that  we  only use  topological information  to
compute the uncertainty.   Thus, we
deliberately neglect  the active role that  a given node  can play as,
for example,  a processing unit, or the  different
weights  of the  paths.  However, it is worth to mention that entropy can  be generalized for DAGs where links are weighted by a probability to be chosen in the process of reaching the maximal.


\section{Limit cases: maximum and minimum uncertainty}
\label{Extremal}
Let us illustrate our previous results by exploring two limit cases, namely DAGs having zero or maximal uncertainty.
In  this  section  we  identify  those  feed-forward  structures  which,
containing $n$ nodes  and without a predefined  number  of links,
minimize   or   maximize  the   above   uncertainties. In this way, for example, a chain having $m=1$ will display
$H({\cal G})=0$, whereas its somehow opposite graph, the star having $m=n-1$ will have $H({\cal G})=\log(n-1)$. 
The derivation of the limit scenarios will be more sophisticated, due to the active role of combinatorics in defining the paths.
The   minimum
uncertainties are obtained when the  graph ${\cal
  G}$  is  a  special  kind  of  tree,  to  be  described  below. Afterwards, we also derive  the  graph configuration  with  maximum  entropy.   The
conceptual   starting  point   of   this derivation   is  the   graph
representation of the linear order.

\begin{figure}
\begin{center}
\includegraphics[width=8cm]{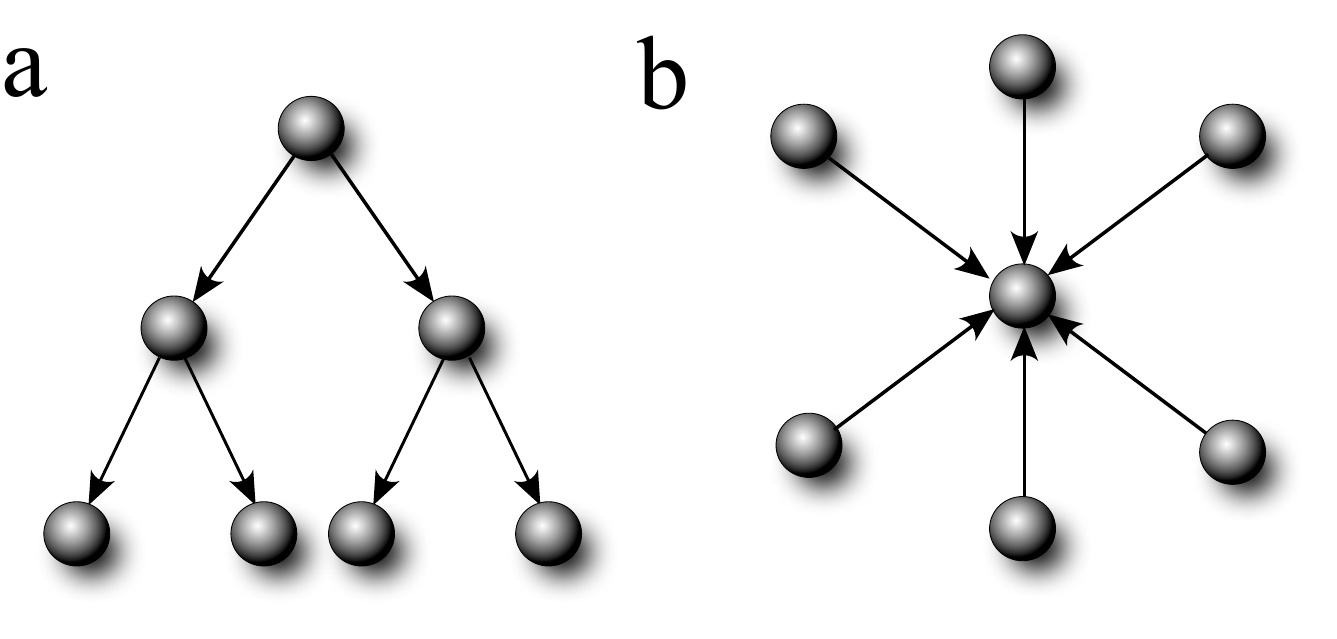}
\caption{A topological reversible structure featured by a tree DAG structure, $H({\cal G})=0$ (a). A topologically irreversible DAG featured by a star DAG with $m=n-1$. Notice that for a star graph $H({\cal G})=\log (n-1)$ where  $n=7$ in this particular case (b).}
\label{fig3}
\end{center}
\end{figure}
\subsection{ Zero Uncertainty: Trees}
\label{Trees}
Imagine a random walker exploring a (directed) tree containing only a
single maximal (fig. \ref{fig3}a). From  such a maximal node,
there exists  only one  path to  a given node. In the evolutionary context, a single ancestor is at the root of all evolutionary tree \cite{Futuyma2006}.   Thus, the  process of
recovering   the  history   of   the  random   walker  up to its initial condition is   completely
deterministic, and no  uncertainty can be associated to  it -in purely
topological terms.   Formally, we  recognize two defining  features on
trees, namely:
\begin{itemize}
\item
$m=1$ 
\item
$(\forall v_i\in V\setminus M)(k_{in}(v_i)=1)$. 
\end{itemize}
We thus conclude that there  is no uncertainty in recovering the flow,
since the two reported properties are enough to conclude that there is
$1$ and  only $1$ path  to go from  $M$ to any $v_i\in  V\setminus M$.
This  agrees   with  the  intuitive   idea  that  trees   are  perfect
hierarchical structures.

This result complements the more standard scenario
 of the forward, downstream scenario paths followed by a random walker on a tree \cite{Schuster2010}.
 It is worth noting that evolutionary trees, particularly in unicellular organisms, have been found to be a poor representation of the actual evolutionary process \cite{Dagan2008,Dagan2009}.

\subsection{Maximum Uncertainty}
\label{MaximalUncertainty}
Now we consider the maximum entropic scenario. For this purpose,
we cut the problem in  two pieces: First, we constructively obtain the
feed forward  graph containing  $m$ maximal nodes  maximizing $H({\cal
  G})$.  Once we identified such  a feed forward configuration, we ask
for the $m$ that maximizes such a quantity.

\subsubsection{The linear ordering in $V\setminus M$.}
\label{LinearOrdering}
Let ${\cal G}$ be a feed-forward organized graph containing $n$ nodes,
where $m$ of them are  maximal.  Since for the entropy computation all
nodes  become indistinguishable,  let  $g(m, n)$  be  the ensemble  of
different possible  feed-forward configurations containing  $n$ nodes,
where $m$  of them  are maximal.  We  are looking  for a graph,  to be
written  as  $\tilde{\cal  G}\in  g(n,m)$, such  that  $\forall  {\cal
  G}_i\in g(m, n) $:
\begin{equation}
{\cal G}_i\subseteq \tilde{\cal G},
\end{equation}
i.e., a graph containing all  possible links, preserving the number of
maximal nodes.   This implies, as defined  in section \ref{BasicGDef},
eq. (\ref{Linear}), that  we must add links to  the set $V\setminus M$
until  it becomes  {\em  linearly ordered},  attending to  a
labeling  of  nodes  which   respect  the  ordering  depicted  by  the
feed-forward graph (see fig. \ref{fig1}c).  Once  we have the
set of nodes $V\setminus M$ linearly ordered, we proceed to generate a
link from any  node $v_i\in M$ to any node  $v_k\in V\setminus M$.  We
thus obtain a feed forward graph containing $m$ maximal nodes and only
$1$  minimal node.   In  the  above constructed  graph,  any new  link
creates a  cycle or destroys  a maximal vertex. Furthermore,  given two
fixed values of $m$ and $n$, it is straightforward to demonstrate that
it maximizes any entropy based on paths: Any feed-forward graph of the
ensemble $g(m, n)$ other than $\tilde{\cal G}$ is obtained by removing
edges of $\tilde{\cal G}$.  This edge removal process will necessarily result in a
reduction of uncertainty.

For  the sake  of clarity  we differentiate  the labeling  of  $M$ and
$V\setminus M$ when working with $\tilde{\cal G}$. Specifically, nodes
$v_i\in V\setminus M$  will be labeled sequentially from  $1$ to $n-m$
respecting  the ordering  defined in  eq. (\ref{i>j}).   This labeling
will  be widely  used  in the  forthcoming  sections. Furthermore,  we
recall that no special labeling other than different natural numbers is
needed for $v_k\in M$, since there will be no ambiguous
situations.   Given  the labeling  proposed
above, and starting from eq. (\ref{Pivi}) the number of paths in  $\tilde{\cal G}$
from $M$ to $v_i\in V\setminus M$ will be:
\begin{eqnarray}
|\Pi(v_i)|&=&\sum_{j\leq L({\cal G})}\sum_{l:v_l\in M}(\mathbf{A}^T({\cal G}))^j_{il}\nonumber\\
&=&\sum_{l:v_l\in M}\sum_{j\leq i}{i \choose j}\nonumber\\
&=&m\left( \sum_{j\leq i}{i \choose j}\right)=m\cdot 2^{i-1}.
\label{TotalPaths}
\end{eqnarray}

\subsubsection{The explicit form of entropies in the linear ordering of $V\setminus M$.}
\label{entropiesLinear}

We first bound $H(\tilde{\cal  G})$ using Jensen's inequality. Indeed,
from eq. (\ref{Jensen})  we can derive an upper  bound for $H(\tilde{\cal
  G})$, namely
\begin{eqnarray}
H(\tilde{\cal G})\leq \log m+\frac{\log 2}{2}(n-m-1).
\end{eqnarray}
We  can go  further, first  computing the  probabilities  defining the
matrix $\Phi({\tilde{\cal G}})$.  To compute these probabilities, let
us suppose we are in node $v_i\in V\setminus M$. The first observation
is that the probability to reach one maximal is
$\frac{1}{m}$.
What about $v_1$, i.e., the first  node we find after the maximal set?
We  observe that,  from the  node $v_i$,  the situation  is completely
analogous to the situation where  there are $m+1$ maximal nodes, since
the probability to pass through  $v_1$ does not depend on what happens
{\em above} $v_1$. Therefore:
\begin{equation}
\phi_{i1}=\frac{1}{m+1}.\nonumber
\end{equation}
and running the reasoning from $v_1$ to $v_{i-1}$, we find that:
\begin{equation}
\phi_{ik}=\frac{1}{m+k}\;\; (k<i).\nonumber
\label{phik}
\end{equation}
Interestingly,  for $k<i$,  $\phi_{ik}$  is invariant,  no matter  the
value  of $i$.   This leads  matrix $\Phi(\tilde{\cal  G})$ to  be:
\begin{equation}
\Phi(\tilde{\cal G})=\left(
\begin{array}{lllll}
1&0&0&...&0\\
\frac{1}{m+1}&1&0&...&0\\
\frac{1}{m+1}&\frac{1}{m+2}&1&...&0\\
...&&&&...\\
...&&&&...\\
\frac{1}{m+1}&\frac{1}{m+2}&\frac{1}{m+3}&..&1
\end{array}
\right),
\label{PhiMax}
\end{equation}
and the  final expression is obtained  by observing that
\begin{equation}
h_L(v_k)=\log(m+k-1),\nonumber
\label{hLMaxk} 
\end{equation}
and therefore,  inserting  it  and (\ref{PhiMax})  into  eq.  (\ref{HFHL}),
we obtain after some algebra:
\begin{equation}
H(\tilde{\cal G})=\frac{n}{n-m}\sum_{i\leq n-m}f(v_i),
\label{ExplicitLinear}
\end{equation}
where $f(v_i)$ is a  function $f:V\setminus M\to\mathbb{R}^+$,
\begin{equation}
f(v_i)=\frac{\log(m+i-1)}{m+i}.
\label{f(v_i)}
\end{equation}
We  can see that  the value entropy is  reduced  to the
computation of the average of $f$ over the set $V\setminus M$. If $\tilde{\cal G}$ contains $n$ nodes, being $m$ of them the maximal ones
we will refer to this average as $\langle f(n,m)\rangle$, defined as:
\begin{equation}
\langle f(n,m)\rangle=\frac{1}{n-m}\sum_{i\leq n-m}f(v_i)
\label{f(n,m)}
\end{equation}

\subsubsection{Absolute maxima of entropies}
\label{AbsoluteMaximums}

What  is  the  relation  between  $n$ and  $m$  maximizing  the  above
entropies?  As we  shall see, given a fixed value  of $n$, the absolute
maximum is found in the  linear ordering above defined at $m^*=2$, for
graphs sizes $n\gg 1 $. 
To support the above claim, let us first notice that:
\begin{equation}
 \left. Q(\tilde{\cal G})\right|_{m=2}=\left.Q(\tilde{\cal G})\right|_{m=1},\nonumber
\end{equation}
enabling us to  derive  the  first
inequality:
\begin{eqnarray}
\left.H(\tilde{\cal G})\right|_{m=2}&=&\frac{1}{n-2}\left.Q(\tilde{\cal G})\right|_{m=1}\nonumber\\
&>&\frac{1}{n-1}\left. Q(\tilde{\cal G})\right|_{m=1}\nonumber\\
&=&\left.H(\tilde{\cal G})\right|_{m=1}.
\label{ineqM1}
\end{eqnarray}
Once       we       demonstrated       that       $\left.H(\tilde{\cal
  G})\right|_{m=2}>\left.H(\tilde{\cal  G})\right|_{m=1}$,  we proceed
to           demonstrate           that           $\left.H(\tilde{\cal
  G})\right|_{m=2}>\left.H(\tilde{\cal   G})\right|_{m=3}$.   To  this
end,  let  us  first  observe  a  key  property  of  $f$,  defined  in
eq.    (\ref{f(v_i)}).    Indeed,    we   observe    that    $(\forall
\epsilon>0)(\exists k_{\epsilon}):(\forall k>k_{\epsilon})$,
\begin{equation}
f(v_k)<\epsilon,
\label{epsilon}
\end{equation} 
provided that $n$ is large enough.  From this property, and since $\langle f(n,m)\rangle$ is an average -see eq. (\ref{f(n,m)})- we can be sure
that $(\exists n^*):(\forall n>n^*)$,
\begin{equation}
\frac{\log 2}{3}>\langle f (n,3)\rangle,
\end{equation}
by choosing appropriately $n$ in such  a way that we have enough terms
lower  than a  given $\epsilon$  to obtain  the above  desired result.
Thus, from eq. (\ref{f(n,m)}) and knowing that
\begin{eqnarray}
\left. H(\tilde{\cal G})\right|_{m=2}-\left.H(\tilde{\cal G})\right|_{m=3}\propto\frac{\log 2}{3}-
\langle f (n,3)\rangle,\nonumber
\end{eqnarray}
(with proportionally factor equal to $n/(n-2)$)
we can conclude that
\begin{equation}
\left.H(\tilde{\cal G})\right|_{m=2}>\left.H(\tilde{\cal G})\right|_{m=3}.\nonumber
\end{equation}
The general case easily derives from the same reasoning, since:
\begin{eqnarray}
\left.H(\tilde{\cal  G})\right|_{m=k}-\left.H(\tilde{\cal  G})\right|_{m=k+1}&\propto& \frac{\log(k+1)}{k}-\langle f (n,k+1)\rangle,\nonumber
\end{eqnarray}
and thus, we can conclude that:
\begin{equation}
(\forall k\leq 2)\;\;\left.H(\tilde{\cal  G})\right|_{m=k}>\left.H(\tilde{\cal  G})\right|_{m=k+1}.
\end{equation}
This closes  the demonstration that $\tilde{\cal  G}$ containing $m=2$
is the  most entropic graph provided that $n>14$, according to numerical computations.


\section{Discussion}
\label{HierarchyIrreversibility}

In this  paper we address the problem of quantifying path dependencies using the DAG metaphor. 
To this goal, we introduce the concept  of topological  reversibility  as a
fundamental feature of causal processes  that can be depicted by a DAG
structure. The intuitive definition  is rather simple: A system formed
by an  aggregation of causal processes is  topologically reversible if
we can recover all causal paths  with no other information than the one
provided by the graph topology. If graph topology induces some kind of
ambiguity  in  the  backward  process,   the graph  is  said  to  be
topologically  irreversible, and additional  information is  needed to
build the backward  flows. 

We  provided  the analytical  form of  the
uncertainty (the  amount of extra  information needed) arising  in the
reversion process by  uncoupling the combinatorial information encoded
by the graph  structure from the contributions  of the local
connectivity  patterns  of  individual   nodes,  as  depicted  in  eqs.
(\ref{HFHL}, \ref{Q}).  It  is worth noting that  all our results are derived from just two basic concepts: The adjacency matrix of the graph
and the  definition of entropy.  Furthermore, we  offer a constructive
derivation of the two limit cases, namely trees (as the reversible
ones), and linear  ordered graphs (having two maximal  nodes) as  the most uncertain  ones.

According to our results, only a tree DAG is topologically reversible. However, beyond this singular case, the quantification of topological  irreversibility by using the entropy proposed here could provide insights in the characterization of feed forward systems. An illustrative case-study can be found  precisely in biological evolution. The standard view of the tree of life involves a directional, upward time-arrow where the genetic structure of a given species (its genome) derives from some ancestor after splitting (speciation) events. One would think that this classical but  too simplistic view of evolution as a tree gives  a topologically  reversible lineage of genes, changing by mutations and passing from the initial ancestor to current species in a vertical inheritance. However, it has been recently evidenced that the so-called horizontal gene transfer among unrelated species may have had a deep impact in the evolution and diversification in microbes \cite{Dagan2009}. According to this genetic mechanism the tree-like and thus the logical/topological reversibility is broken by the presence of cross-links between \textit{brother} species. At the light of these evidences, tree-based phylogenies become unrealistic. In this context, our theoretical approach provides a suitable framework for the characterization of the logical irreversibility of biological evolution and, in general, for any process where time or energy dissipation impose a feed-forward chart of events.  Further research in this topic will contribute to understand the causal structure of evolutionary processes.

\begin{acknowledgments}
This  work  was  supported   by  the  EU  $6^{th}$  framework  project
ComplexDis (NEST-043241, CRC and JG), the UTE project CIMA (JG), James
McDonnell Foundation  (BCM and RVS)  and the Santa Fe Institute  (RVS). We
thank Ivan Bezdomny and Complex System Lab members for fruitful conversations.
\end{acknowledgments}




\begin{thebibliography}{38}
\expandafter\ifx\csname natexlab\endcsname\relax\def\natexlab#1{#1}\fi
\expandafter\ifx\csname bibnamefont\endcsname\relax
  \def\bibnamefont#1{#1}\fi
\expandafter\ifx\csname bibfnamefont\endcsname\relax
  \def\bibfnamefont#1{#1}\fi
\expandafter\ifx\csname citenamefont\endcsname\relax
  \def\citenamefont#1{#1}\fi
\expandafter\ifx\csname url\endcsname\relax
  \def\url#1{\texttt{#1}}\fi
\expandafter\ifx\csname urlprefix\endcsname\relax\def\urlprefix{URL }\fi
\providecommand{\bibinfo}[2]{#2}
\providecommand{\eprint}[2][]{\url{#2}}

\bibitem[{\citenamefont{Csardi et~al.}(2007)\citenamefont{Csardi, Strandburg,
  Zalanyi, Tobochnik, and Erdi}}]{Csardi2007}
\bibinfo{author}{\bibfnamefont{G.}~\bibnamefont{Csardi}},
  \bibinfo{author}{\bibfnamefont{K.~J.} \bibnamefont{Strandburg}},
  \bibinfo{author}{\bibfnamefont{L.}~\bibnamefont{Zalanyi}},
  \bibinfo{author}{\bibfnamefont{J.}~\bibnamefont{Tobochnik}},
  \bibnamefont{and} \bibinfo{author}{\bibfnamefont{P.}~\bibnamefont{Erdi}},
  \bibinfo{journal}{Physica A} \textbf{\bibinfo{volume}{374}},
  \bibinfo{pages}{783} (\bibinfo{year}{2007}).

\bibitem[{\citenamefont{Karrer and Newman}(2009)}]{Karrer2009}
\bibinfo{author}{\bibfnamefont{B.}~\bibnamefont{Karrer}} \bibnamefont{and}
  \bibinfo{author}{\bibfnamefont{M.~E.~J.} \bibnamefont{Newman}},
  \bibinfo{journal}{Phys Rev Lett} \textbf{\bibinfo{volume}{102}},
  \bibinfo{pages}{128701} (\bibinfo{year}{2009}).

\bibitem[{\citenamefont{Lehmann et~al.}(2003)\citenamefont{Lehmann, Lautrup,
  and Jackson}}]{Lehmann:2003}
\bibinfo{author}{\bibfnamefont{S.}~\bibnamefont{Lehmann}},
  \bibinfo{author}{\bibfnamefont{B.}~\bibnamefont{Lautrup}}, \bibnamefont{and}
  \bibinfo{author}{\bibfnamefont{A.~D.} \bibnamefont{Jackson}},
  \bibinfo{journal}{Phys. Rev. E} \textbf{\bibinfo{volume}{68}},
  \bibinfo{pages}{026113} (\bibinfo{year}{2003}).

\bibitem[{\citenamefont{Valverde et~al.}(2007)\citenamefont{Valverde, Sol\'e,
  Bedau, and Packard}}]{Valverde2007}
\bibinfo{author}{\bibfnamefont{S.}~\bibnamefont{Valverde}},
  \bibinfo{author}{\bibfnamefont{R.~V.} \bibnamefont{Sol\'e}},
  \bibinfo{author}{\bibfnamefont{M.~A.} \bibnamefont{Bedau}}, \bibnamefont{and}
  \bibinfo{author}{\bibfnamefont{N.}~\bibnamefont{Packard}},
  \bibinfo{journal}{Phys Rev E Stat Nonlin Soft Matter Phys}
  \textbf{\bibinfo{volume}{76}}, \bibinfo{pages}{056118}
  (\bibinfo{year}{2007}).

\bibitem[{\citenamefont{Clay}(1971)}]{Clay1971}
\bibinfo{author}{\bibfnamefont{R.}~\bibnamefont{Clay}},
  \emph{\bibinfo{title}{Nonlinear networks and systems}}
  (\bibinfo{publisher}{John Wiley \& Sons Inc, New York},
  \bibinfo{year}{1971}).

\bibitem[{\citenamefont{Haykin}(1999)}]{Haykin1999}
\bibinfo{author}{\bibfnamefont{S.}~\bibnamefont{Haykin}},
  \emph{\bibinfo{title}{Neural Networks : a Comprehensive Foundation}}
  (\bibinfo{publisher}{Prentice-Hall. London}, \bibinfo{year}{1999}).

\bibitem[{\citenamefont{Frank and Frisch}(1971)}]{Frank1972}
\bibinfo{author}{\bibfnamefont{H.}~\bibnamefont{Frank}} \bibnamefont{and}
  \bibinfo{author}{\bibfnamefont{I.~T.} \bibnamefont{Frisch}},
  \emph{\bibinfo{title}{Communication, transmission and transportation
  networks}} (\bibinfo{publisher}{Addison-Wesley (Reading Mass)},
  \bibinfo{year}{1971}).

\bibitem[{\citenamefont{Rodr\'iguez-Iturbe and
  Rinaldo}(1997)}]{Rodriguez-Iturbe1997}
\bibinfo{author}{\bibfnamefont{I.}~\bibnamefont{Rodr\'iguez-Iturbe}}
  \bibnamefont{and} \bibinfo{author}{\bibfnamefont{A.}~\bibnamefont{Rinaldo}},
  \emph{\bibinfo{title}{Fractal River Basins. Chance and Self-organization}}
  (\bibinfo{publisher}{Cambridge University Press. Cambridge.},
  \bibinfo{year}{1997}).

\bibitem[{\citenamefont{Bonchev and Rouvray}(2005)}]{Bonchev2005}
\bibinfo{author}{\bibfnamefont{D.}~\bibnamefont{Bonchev}} \bibnamefont{and}
  \bibinfo{author}{\bibfnamefont{D.~H.} \bibnamefont{Rouvray}},
  \emph{\bibinfo{title}{Complexity in Chemistry, Biology, and Ecology}}
  (\bibinfo{publisher}{Springer, New York.}, \bibinfo{year}{2005}).

\bibitem[{\citenamefont{Bennett}(1973)}]{Bennett:1973}
\bibinfo{author}{\bibfnamefont{C.~H.} \bibnamefont{Bennett}},
  \bibinfo{journal}{IBM J. Res. Dev.} \textbf{\bibinfo{volume}{17}},
  \bibinfo{pages}{525} (\bibinfo{year}{1973}).

\bibitem[{\citenamefont{Landauer}(1961)}]{Landauer:1961}
\bibinfo{author}{\bibfnamefont{R.}~\bibnamefont{Landauer}},
  \bibinfo{journal}{IBM Journal of Research and Development}
  \textbf{\bibinfo{volume}{5}}, \bibinfo{pages}{183} (\bibinfo{year}{1961}).

\bibitem[{\citenamefont{Fontana and Buss}(1994)}]{Fontana1994}
\bibinfo{author}{\bibfnamefont{W.}~\bibnamefont{Fontana}} \bibnamefont{and}
  \bibinfo{author}{\bibfnamefont{L.~W.} \bibnamefont{Buss}},
  \bibinfo{journal}{Proc Natl Acad Sci U S A} \textbf{\bibinfo{volume}{91}},
  \bibinfo{pages}{757} (\bibinfo{year}{1994}).

\bibitem[{\citenamefont{Gould}(1990)}]{Gould1990}
\bibinfo{author}{\bibfnamefont{S.~J.} \bibnamefont{Gould}},
  \emph{\bibinfo{title}{Wonderful Life: The Burgess Shale and the Nature of
  History}} (\bibinfo{publisher}{W. W. Norton \& Company. New York},
  \bibinfo{year}{1990}).

\bibitem[{\citenamefont{de~Groot and Mazur}(1962)}]{DeGroot:1963}
\bibinfo{author}{\bibfnamefont{S.~R.} \bibnamefont{de~Groot}} \bibnamefont{and}
  \bibinfo{author}{\bibfnamefont{P.}~\bibnamefont{Mazur}},
  \emph{\bibinfo{title}{Non-Equilibrium Thermodynamics}}
  (\bibinfo{publisher}{North-Holland. Amsterdam}, \bibinfo{year}{1962}).

\bibitem[{\citenamefont{Lebon et~al.}(2008)\citenamefont{Lebon, Jou, and
  Casas-V\'azquez}}]{Lebon:2008}
\bibinfo{author}{\bibfnamefont{G.}~\bibnamefont{Lebon}},
  \bibinfo{author}{\bibfnamefont{D.}~\bibnamefont{Jou}}, \bibnamefont{and}
  \bibinfo{author}{\bibfnamefont{J.}~\bibnamefont{Casas-V\'azquez}},
  \emph{\bibinfo{title}{Understanding Nonequilibrium Thermodynamics}}
  (\bibinfo{publisher}{Springer, Berlin, 2008}, \bibinfo{year}{2008}).

\bibitem[{\citenamefont{Schuster}(2010)}]{Schuster2010}
\bibinfo{author}{\bibfnamefont{P.}~\bibnamefont{Schuster}},
  \bibinfo{journal}{Complexity} \textbf{\bibinfo{volume}{In press}}
  (\bibinfo{year}{2010}).

\bibitem[{\citenamefont{Anand and Bianconi}(2009)}]{Bianconi:2009}
\bibinfo{author}{\bibfnamefont{K.}~\bibnamefont{Anand}} \bibnamefont{and}
  \bibinfo{author}{\bibfnamefont{G.}~\bibnamefont{Bianconi}},
  \bibinfo{journal}{Phys. Rev. E} \textbf{\bibinfo{volume}{80}},
  \bibinfo{pages}{045102} (\bibinfo{year}{2009}).

\bibitem[{\citenamefont{Dehmer}(2008)}]{Dehmer:2008b}
\bibinfo{author}{\bibfnamefont{M.}~\bibnamefont{Dehmer}},
  \bibinfo{journal}{Appl. Artif. Intell.} \textbf{\bibinfo{volume}{22}},
  \bibinfo{pages}{684} (\bibinfo{year}{2008}), ISSN \bibinfo{issn}{0883-9514}.

\bibitem[{\citenamefont{Dehmer et~al.}(2008)\citenamefont{Dehmer, Borgert, and
  Emmert-Streib}}]{Dehmer:2008}
\bibinfo{author}{\bibfnamefont{M.}~\bibnamefont{Dehmer}},
  \bibinfo{author}{\bibfnamefont{S.}~\bibnamefont{Borgert}}, \bibnamefont{and}
  \bibinfo{author}{\bibfnamefont{F.}~\bibnamefont{Emmert-Streib}},
  \bibinfo{journal}{PLoS ONE} \textbf{\bibinfo{volume}{3}},
  \bibinfo{pages}{e3079} (\bibinfo{year}{2008}).

\bibitem[{\citenamefont{Estrada}(2009)}]{Estrada:2009}
\bibinfo{author}{\bibfnamefont{E.}~\bibnamefont{Estrada}},
  \bibinfo{journal}{Phys. Rev. E} \textbf{\bibinfo{volume}{80}},
  \bibinfo{pages}{026104} (\bibinfo{year}{2009}).

\bibitem[{\citenamefont{Schneidman et~al.}(2003)\citenamefont{Schneidman,
  Still, Berry, and Bialek}}]{Schneidman:2003}
\bibinfo{author}{\bibfnamefont{E.}~\bibnamefont{Schneidman}},
  \bibinfo{author}{\bibfnamefont{S.}~\bibnamefont{Still}},
  \bibinfo{author}{\bibfnamefont{M.~J.} \bibnamefont{Berry}}, \bibnamefont{and}
  \bibinfo{author}{\bibfnamefont{W.}~\bibnamefont{Bialek}},
  \bibinfo{journal}{Phys. Rev. Lett.} \textbf{\bibinfo{volume}{91}},
  \bibinfo{pages}{238701} (\bibinfo{year}{2003}).

\bibitem[{\citenamefont{Sol\'e and Valverde}(2004)}]{Sole:2004}
\bibinfo{author}{\bibfnamefont{R.~V.} \bibnamefont{Sol\'e}} \bibnamefont{and}
  \bibinfo{author}{\bibfnamefont{S.}~\bibnamefont{Valverde}}, in
  \emph{\bibinfo{booktitle}{Networks: Structure, Dynamics and Function, Lecture
  Notes in Physics.}} (\bibinfo{publisher}{Springer-Verlag},
  \bibinfo{year}{2004}), pp. \bibinfo{pages}{189--210}.

\bibitem[{\citenamefont{Ulanowicz}(1986)}]{Ulanowicz1986}
\bibinfo{author}{\bibfnamefont{R.~E.} \bibnamefont{Ulanowicz}},
  \emph{\bibinfo{title}{Growth and Development: Ecosystems Phenomenology.}}
  (\bibinfo{publisher}{Springer, New York.}, \bibinfo{year}{1986}).

\bibitem[{\citenamefont{Bollob\'as}(1998)}]{Bollobas:1998}
\bibinfo{author}{\bibfnamefont{B.}~\bibnamefont{Bollob\'as}},
  \emph{\bibinfo{title}{Modern Graph Theory}} (\bibinfo{publisher}{Springer},
  \bibinfo{year}{1998}), \bibinfo{edition}{corrected} ed., ISBN
  \bibinfo{isbn}{0387984887}.

\bibitem[{\citenamefont{Gross and Yellen}(1998)}]{Gross:1998}
\bibinfo{author}{\bibfnamefont{J.}~\bibnamefont{Gross}} \bibnamefont{and}
  \bibinfo{author}{\bibfnamefont{J.}~\bibnamefont{Yellen}},
  \emph{\bibinfo{title}{Graph Theory and its applications}}
  (\bibinfo{publisher}{CRC, Boca Raton, Florida}, \bibinfo{year}{1998}).

\bibitem[{\citenamefont{Kelley}(1955)}]{Kelley:1955}
\bibinfo{author}{\bibfnamefont{J.}~\bibnamefont{Kelley}},
  \emph{\bibinfo{title}{General Topology}}, Graduate Texts in Mathematics, 27,
  1975 (\bibinfo{publisher}{Van Nostrand}, \bibinfo{year}{1955}).

\bibitem[{\citenamefont{Suppes}(1960)}]{Suppes:1960}
\bibinfo{author}{\bibfnamefont{P.}~\bibnamefont{Suppes}},
  \emph{\bibinfo{title}{Axiomatic Set Theory}} (\bibinfo{publisher}{Dover. New
  York}, \bibinfo{year}{1960}).

\bibitem[{\citenamefont{Ash}(1990)}]{Ash:1990}
\bibinfo{author}{\bibfnamefont{R.~B.} \bibnamefont{Ash}},
  \emph{\bibinfo{title}{Information Theory}} (\bibinfo{publisher}{New York.
  Dover}, \bibinfo{year}{1990}).

\bibitem[{\citenamefont{Cover and Thomas}(1991)}]{Thomas:2001}
\bibinfo{author}{\bibfnamefont{T.~M.} \bibnamefont{Cover}} \bibnamefont{and}
  \bibinfo{author}{\bibfnamefont{J.~A.} \bibnamefont{Thomas}},
  \emph{\bibinfo{title}{Elements of Information Theory}}
  (\bibinfo{publisher}{John Wiley and Sons. New York}, \bibinfo{year}{1991}).

\bibitem[{\citenamefont{Khinchin}(1957)}]{Khinchin:1957}
\bibinfo{author}{\bibfnamefont{A.~I.} \bibnamefont{Khinchin}},
  \emph{\bibinfo{title}{Mathematical Foundations of Information Theory}}
  (\bibinfo{publisher}{Dover, New York}, \bibinfo{year}{1957}).

\bibitem[{\citenamefont{Shannon}(1948)}]{Shannon:1948}
\bibinfo{author}{\bibfnamefont{C.~E.} \bibnamefont{Shannon}},
  \bibinfo{journal}{Bell System Technical Journal}
  \textbf{\bibinfo{volume}{27}}, \bibinfo{pages}{379} (\bibinfo{year}{1948}).

\bibitem[{\citenamefont{Lagomarsino et~al.}(2006)\citenamefont{Lagomarsino,
  Jona, and Bassetti}}]{Lagomarsino2006}
\bibinfo{author}{\bibfnamefont{M.~C.} \bibnamefont{Lagomarsino}},
  \bibinfo{author}{\bibfnamefont{P.}~\bibnamefont{Jona}}, \bibnamefont{and}
  \bibinfo{author}{\bibfnamefont{B.}~\bibnamefont{Bassetti}}, in
  \emph{\bibinfo{booktitle}{CMSB}} (\bibinfo{year}{2006}), pp.
  \bibinfo{pages}{227--241}.

\bibitem[{\citenamefont{Rodr\'iguez-Caso
  et~al.}(2009)\citenamefont{Rodr\'iguez-Caso, Corominas-Murtra, and
  Sol\'e}}]{Rodriguez-Caso2009}
\bibinfo{author}{\bibfnamefont{C.}~\bibnamefont{Rodr\'iguez-Caso}},
  \bibinfo{author}{\bibfnamefont{B.}~\bibnamefont{Corominas-Murtra}},
  \bibnamefont{and} \bibinfo{author}{\bibfnamefont{R.~V.}
  \bibnamefont{Sol\'e}}, \bibinfo{journal}{Mol Biosyst}
  \textbf{\bibinfo{volume}{5}}, \bibinfo{pages}{1617} (\bibinfo{year}{2009}).

\bibitem[{\citenamefont{Jensen}(2002)}]{Jensen:2001}
\bibinfo{author}{\bibfnamefont{F.~V.} \bibnamefont{Jensen}},
  \emph{\bibinfo{title}{Bayesian Networks and Decision Graphs}}, Information
  Science and Statistics (\bibinfo{publisher}{Springer}, \bibinfo{year}{2002}).

\bibitem[{\citenamefont{Futuyma}(2005)}]{Futuyma2006}
\bibinfo{author}{\bibfnamefont{D.~J.} \bibnamefont{Futuyma}},
  \emph{\bibinfo{title}{Evolution}} (\bibinfo{publisher}{Sinauer Associates.
  Sunderland}, \bibinfo{year}{2005}).

\bibitem[{\citenamefont{Dagan et~al.}(2008)\citenamefont{Dagan, Artzy-Randrup,
  and Martin}}]{Dagan2008}
\bibinfo{author}{\bibfnamefont{T.}~\bibnamefont{Dagan}},
  \bibinfo{author}{\bibfnamefont{Y.}~\bibnamefont{Artzy-Randrup}},
  \bibnamefont{and} \bibinfo{author}{\bibfnamefont{W.}~\bibnamefont{Martin}},
  \bibinfo{journal}{Proc Natl Acad Sci U S A} \textbf{\bibinfo{volume}{105}},
  \bibinfo{pages}{10039} (\bibinfo{year}{2008}).

\bibitem[{\citenamefont{Dagan and Martin}(2009)}]{Dagan2009}
\bibinfo{author}{\bibfnamefont{T.}~\bibnamefont{Dagan}} \bibnamefont{and}
  \bibinfo{author}{\bibfnamefont{W.}~\bibnamefont{Martin}},
  \bibinfo{journal}{Philos Trans R Soc Lond B Biol Sci}
  \textbf{\bibinfo{volume}{364}}, \bibinfo{pages}{2187} (\bibinfo{year}{2009}).

\bibitem[{\citenamefont{Van~Kampen}(2007)}]{VanKampen:2007}
\bibinfo{author}{\bibfnamefont{N.~G.} \bibnamefont{Van~Kampen}},
  \emph{\bibinfo{title}{Stochastic Processes in Physics and Chemistry, Third
  Edition (North-Holland Personal Library)}} (\bibinfo{publisher}{North
  Holland}, \bibinfo{year}{2007}), \bibinfo{edition}{3rd} ed.

\end{thebibliography}

\end{document}